\documentclass[aip,apl,amsmath,amssymb,reprint]{revtex4-2}

\usepackage{graphicx}
\usepackage{booktabs}

\usepackage{tabularx}
\usepackage{dcolumn}
\usepackage{bm}
\usepackage{mathrsfs} 
\usepackage{color}

\usepackage{lineno}

\usepackage{siunitx}

\usepackage{xcolor}

\newcommand{\sigmape}{$\sigma_{\mathrm{p.e.}}$}

\AtBeginDocument{
\heavyrulewidth=.08em
\lightrulewidth=.05em
\cmidrulewidth=.03em
\belowrulesep=.65ex
\belowbottomsep=0pt
\aboverulesep=.4ex
\abovetopsep=0pt
\cmidrulesep=\doublerulesep
\cmidrulekern=.5em
\defaultaddspace=.5em
}
\begin{document}

\preprint{AIP/123-QED}

\title{Photoelectric absorption cross section of silicon near the band gap from room temperature to sub-Kelvin temperature}


\author{C.~Stanford} \email{cstanford@g.harvard.edu}\affiliation{Department of Physics, Stanford University, Stanford, CA 94305 USA}\affiliation{Department of Physics, Harvard University, Cambridge, MA 02138 USA}

\author{M.J.~Wilson} \affiliation{Department of Physics, University of Toronto, Toronto, ON M5S 1A7, Canada} \affiliation{Institut f{\"u}r  Experimentalphysik,  Universit{\"a}t Hamburg, 22761 Hamburg,  Germany}

\author{B.~Cabrera}\email{cabrera@stanford.edu} \affiliation{Department of Physics, Stanford University, Stanford, CA 94305 USA} \affiliation{SLAC National Accelerator Laboratory/Kavli Institute for Particle Astrophysics and Cosmology, 2575 Sand Hill Road, Menlo Park, CA 94025 USA}

\author{M.~Diamond} \affiliation{Department of Physics, University of Toronto, Toronto, ON M5S 1A7, Canada}

\author{N.A.~Kurinsky} \affiliation{Fermi National Accelerator Laboratory, Center for Particle Astrophysics, Batavia, IL 60510 USA} \affiliation{Kavli Institute for Cosmological Physics, University of Chicago, Chicago, IL 60637, USA}

\author{R.A.~Moffatt} \affiliation{Department of Physics, Stanford University, Stanford, CA 94305 USA}

\author{F.~Ponce} \affiliation{Department of Physics, Stanford University, Stanford, CA 94305 USA}

\author{B.~von~Krosigk} \affiliation{Institut  f{\"u}r  Experimentalphysik,  Universit{\"a}t  Hamburg,  22761  Hamburg,  Germany}

\author{B.A.~Young} \affiliation{Department of Physics, Santa Clara University, Santa Clara, CA 95053 USA}

\date{\today}

\begin{abstract}

The use of cryogenic silicon as a detector medium for dark matter searches is gaining popularity. Many of these searches are highly dependent on the value of the photoelectric absorption cross section of silicon at low temperatures, particularly near the silicon band gap energy, where the searches are most sensitive to low mass dark matter candidates. While such cross section data has been lacking from the literature, previous dark matter search experiments have attempted to estimate this parameter by extrapolating it from higher temperature data. However, discrepancies in the high temperature data have led to order-of-magnitude differences in the extrapolations. In this paper, we resolve these discrepancies by using a novel technique to make a direct, low temperature measurement of the photoelectric absorption cross section of silicon at energies near the band gap (1.2—2.8\,eV).
\end{abstract}

\maketitle

\section{Introduction}

The photoelectric absorption cross section (\sigmape) of silicon at low temperatures is an important parameter for modern experiments that 
use cryogenic silicon as a substrate for the direct detection of dark matter~\cite{hvevRun1,hvevRun2,SuperCDMSSoudan2020,DAMIC2019,SENSEI2019}. Several dark matter signal models depend on this parameter. We discuss these models in more detail in \citet{pexsec_vonKrosigk}, but will highlight two of them here. 

First, the hypothesized kinetic mixing of dark photons and Standard Model photons results in an expected interaction between dark photons and electrons, with a cross section given by\cite{Bloch:2016sjj}:
\begin{equation}
    \sigma_{A^\prime}(E_{A^\prime}) =  \frac{\varepsilon^2}{\beta_{A^\prime}}\sigma_{\mathrm{p.e.}}(E_{A^\prime}),
\end{equation}
\noindent where $E_{A^\prime}$ is the dark photon's total energy, $\beta_{A^\prime}=v_{A^\prime}/c$ is the dark photon's relativistic beta factor, and $\varepsilon$ is the kinetic mixing parameter.

Second, the expected cross section for the interaction of axion-like particles (ALPs) with electrons is given by \cite{Pospelov:2008jk, Fu:2017lfc}:

\begin{equation}
    \sigma_a(E_a) = \sigma_{\mathrm{p.e.}}(E_a) \frac{g_{ae}^2}{\beta_a} \frac{3 E_a^2}{16 \pi \,\alpha\, m_e^2 c^4} \left( 1 - \frac{\beta_a^{2/3}}{3} \right),
\end{equation}

\noindent where $E_a$ is the ALP's total energy, $\beta_a=v_a/c$ is its relativistic beta factor, $\alpha$ is the fine structure constant, $m_e$ is the mass of the electron, and $g_{ae}$ is the axioelectric coupling of the ALP to the electrons.

Note that in both (1) and (2), the interaction rate is directly dependent on the value of \sigmape. We performed an exhaustive literature search~\cite{pexseclit0,pexseclit1,pexseclit2,pexseclit3,pexseclit4,pexseclit5,pexseclit6,pexseclit7,pexseclit8} for measurements of \sigmape\ at low energies ($<\SI{10}{\electronvolt}$), where silicon dark matter experiments are most competitive, but found the data to be lacking for the temperature regime of interest ($<\SI{5}{\kelvin}$).
Although previous dark matter search experiments have attempted to account for the temperature dependence of \sigmape\ by extrapolating from the data found in the aforementioned literature search~\cite{hvevRun1,hvevRun2}, discrepancies in the high temperature data have led to low temperature projections that differ by more than an order of magnitude at energies near the silicon band gap.
The result is a dominating uncertainty in any experimental sensitivity curve for dark matter models that depend on \sigmape. This limitation was the motivation for the direct measurement of \sigmape\ at sub-Kelvin temperatures.

\section{Experimental Setup}\label{sec:exp}

To measure \sigmape, we designed an experiment in which a monochromatic light beam was sent through several silicon ``filters" with varying thicknesses. Then, by comparing the relative transmission, a value for \sigmape\ was obtained, free from many of the systematics that would be present in an absolute transmission measurement. 

\begin{figure}[t]
    \centering
    \includegraphics[width=0.5\textwidth]{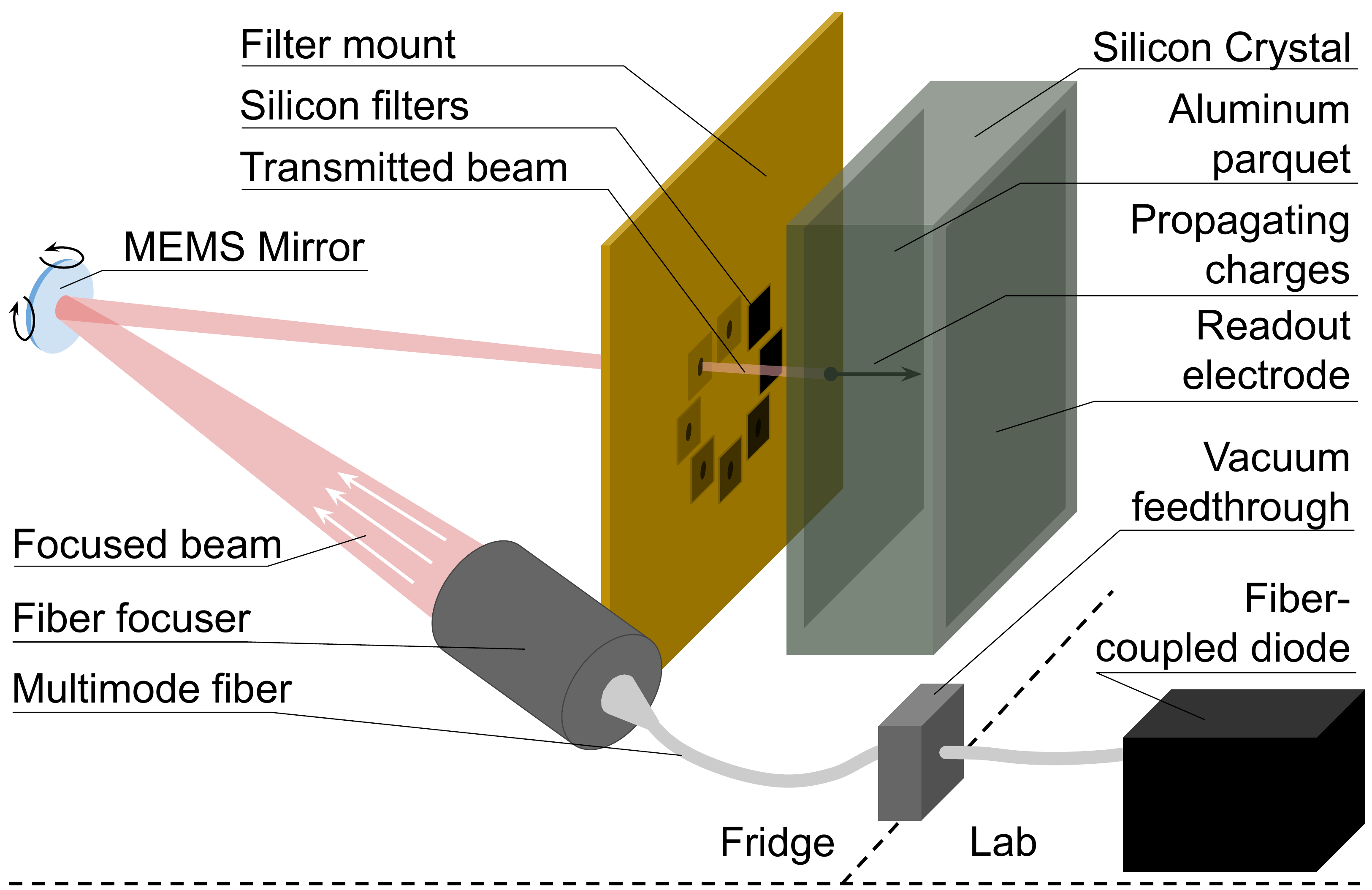}
    \caption{~The detector, which consisted of a \SI{1}{\centi\meter}$\times$\SI{1}{\centi\meter}$\times$\SI{4}{\milli\meter} crystal of high purity silicon, and the laser mechanism, which used a MEMS mirror to scan a focused beam of light pulses across the back face of the crystal in order to produce a 2D image. The crystal was biased via a parquet electrode on the illuminated face.}
    \label{fig:Cartoon}
\end{figure}

The experiment was performed in the same $^3$He cryostat used to previously measure charge propagation in silicon and germanium at low temperatures~\cite{moffatt2019,stanford2019, shank12, shank14}. The cryostat was retrofitted with a new 50/125 multi-mode fiber optic (FO) via a vacuum feed through. The FO was used to illuminate samples at the cold stage with various external light sources (LED/laser diode) of differing wavelengths (see Table~\ref{tab:Diodes}). At the base-temperature stage, the FO was directly coupled to a lens. The beam was focused to a diameter of approximately \SI{200}{\micro\meter} onto a filter mount roughly \SI{150}{mm} away via a 2-axis MEMS mirror, as illustrated in Figure~\ref{fig:Cartoon}. 
The MEMS mirror tilt controls the $x$-$y$ position of the incident beam on the filter mount, with the relaxed state set to the center of the filter mount. Thus, the angle of incidence to any radially symmetric point on the filter mount was the same. 

\begin{figure}[t!]
    \centering
    \includegraphics[width=0.38\textwidth]{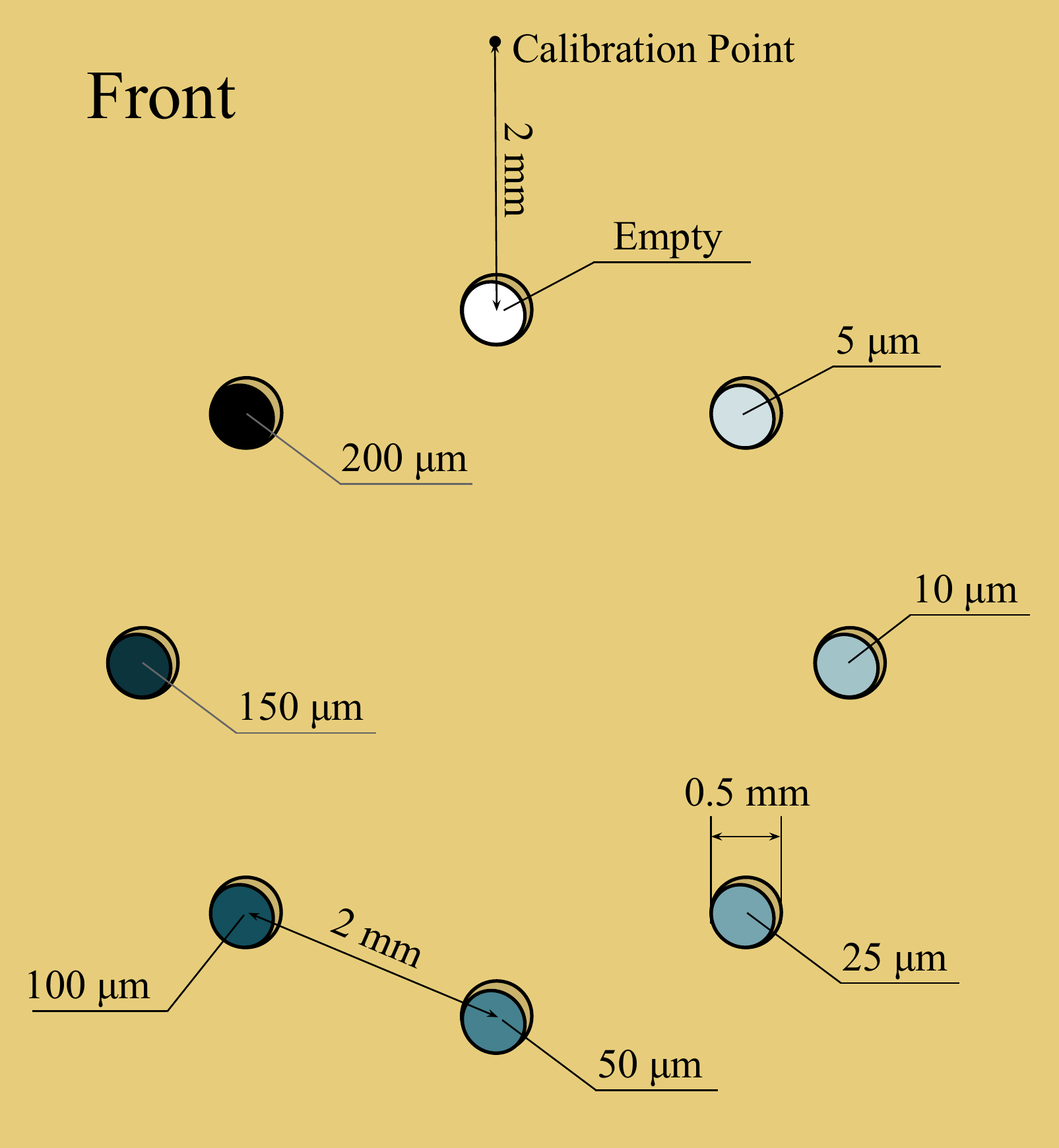}
    \includegraphics[width=0.38\textwidth]{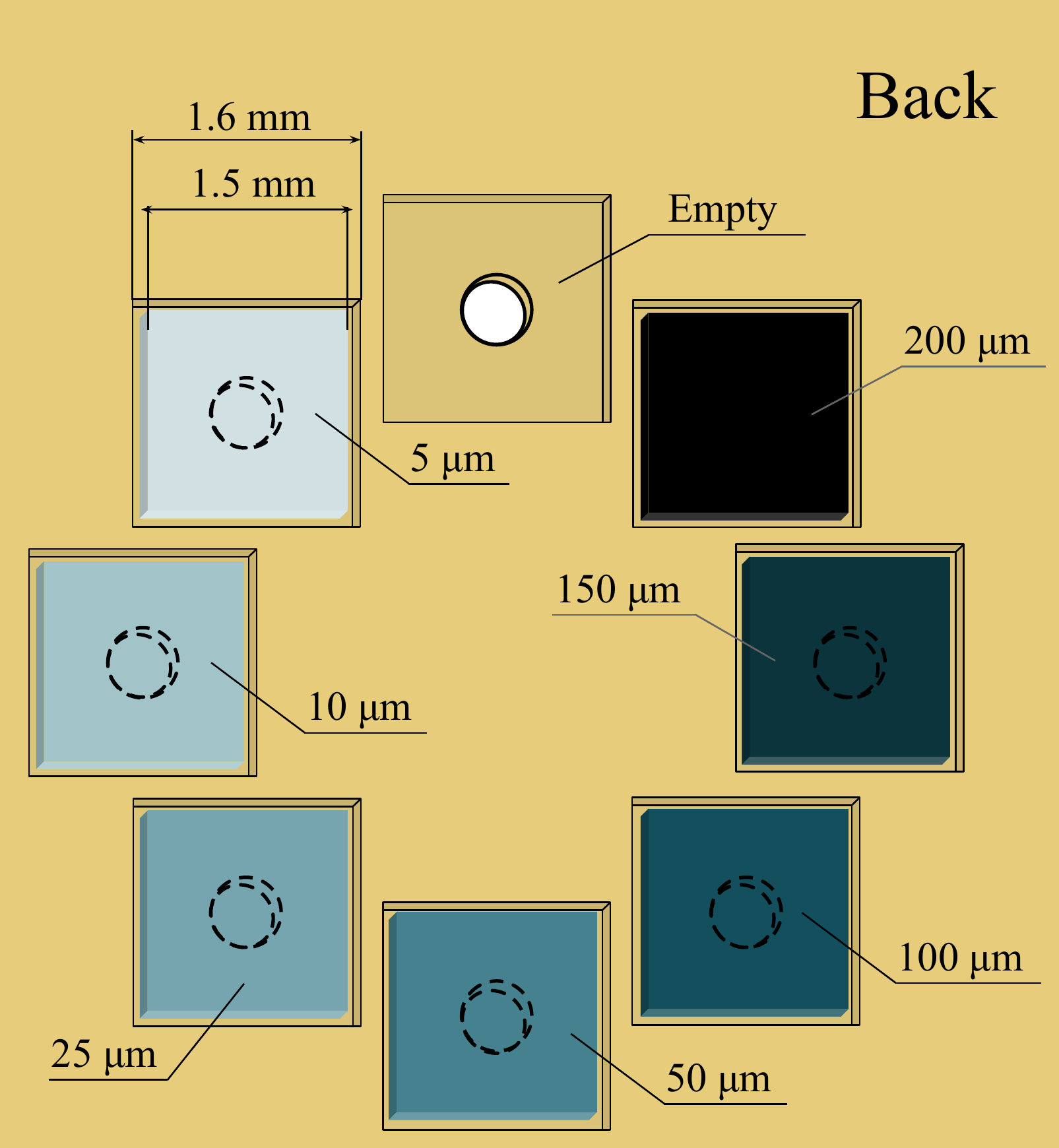}
    \caption{\textbf{Top}:~The front (facing the beam) of the filter mount featured 8 holes arrayed in a circular pattern. When the mirror was at rest, the beam spot fell in the center of the circle. This was done so that the angle of incidence of the beam at each hole (and therefore the amount of reflected light) was the same. At a distance \SI{2}{\milli\meter} radially outward from the empty hole, a calibration point was designated to measure spillover light from the beam spot into neighboring holes. \textbf{Bottom}:~The back (facing the detector crystal) of the filter mount featured square indents aligned with each hole, which held pieces of silicon with different thicknesses. Each piece of silicon was held in place using two small spots of GE varnish at opposing corners.}
    \label{fig:FilterMount}
\end{figure}

The filter mount was made from a \SI{6}{\centi\meter}$^2$ brass plate, illustrated in Figure~\ref{fig:FilterMount}. The front side had eight 0.5 mm diameter through-holes radially symmetric to the mount center. Adjacent holes were \SI{2}{\milli\meter} apart, center-to-center. The back side had 1.6$\times$\SI{1.6}{\milli\meter}$^2$ indents, each centered on a through-hole and used to mount a 1.5$\times$\SI{1.5}{\milli\meter}$^2$ piece of silicon that was held in place with GE varnish. The silicon samples used in this work were cut from seven thinned, boron-doped ({$\rho$} $<$ 100 {$\Omega$}{$\cdot$}cm) CZ Prime $<$100$>$ wafers with thicknesses in \SI{}{\micro\meter} of 5.1$\pm$0.1, 10.0$\pm$0.1, 24.0$\pm$0.2, 49.4$\pm$0.1, 100.2$\pm$0.1, 149.7$\pm$0.2 and 198.8$\pm$0.1. One indent remained empty for calibration purposes.

Approximately \SI{3}{\milli\meter} behind the filter mount (relative to the oncoming photon beam) was a silicon crystal that acted as the detector. This crystal was cut from a \SI{4}{\milli\meter}-thick wafer of undoped ultra-high-purity float-zone silicon ($\sim$15 k$\Omega$-cm). The residual impurity was measured to be p-type with a concentration of $10^{12}$\,cm$^{-3}$. The front (facing the oncoming beam) and back faces of the crystal were \SI{1}{\centi\meter}$\times$\SI{1}{\centi\meter}. The front face was patterned with an aluminum-tungsten mesh electrode, with 20\% coverage~\cite{electrode}, which was used to bias the crystal to \SI{50}{\volt\per\centi\meter}. The back face was covered almost fully with an aluminum thin film that served as a ground electrode. 

The silicon filters were individually illuminated with photons of different wavelengths by manipulating the MEMS mirror. The transmission through the silicon filters was measured as a charge signal in the silicon crystal detector. The charge was collected externally through an amplifier circuit by the data acquisition system (DAQ).

\section{Procedure}\label{sec:Procedure}

The data-taking procedure was as follows. First, the fridge was set to the desired temperature, and a light source was connected to the fiber leading into the fridge. The options for light sources spanned a range of wavelengths, summarized in Table~\ref{tab:Diodes}. Then, a run was performed at that temperature and wavelength. A run involved a scan of the beam spot over each hole, with each scan involving a 14-by-14 grid of individual measurements. To make a single measurement, the diode was pulsed in a train of 32 pulses using a Thorlabs DC2200 LED Driver. The pulses had widths of \SI{20}{\micro\second} (some diodes required longer widths), spaced \SI{8}{\milli\second} apart. These pulses travelled down the fiber, out the focuser, and bounced off the MEMS mirror, which directed them through one of the holes in the filter mount. Then, some of the light from each pulse reflected off the silicon filter present in that hole, while another fraction of the light was absorbed by the filter. The remaining fraction was transmitted through the filter and detected.

\begin{table}
    \centering
    \vspace{0.5cm}
        \begin{tabularx}{9cm}{ *{6}{>{\raggedright\arraybackslash}X}}
            \multicolumn{1}{c}{Peak (nm)}
            &\multicolumn{1}{c}{LHM (nm)}
            &\multicolumn{1}{c}{UHM (nm)}
            &\multicolumn{1}{c}{Peak (eV)}
            &\multicolumn{1}{c}{Type} \\
            \midrule
            \hspace{1em}450                  & \hspace{1em}447                            & \hspace{1em}449                            & \hspace{1em}2.77        & \hspace{1em}Laser        \\
            \hspace{1em}530                  & \hspace{1em}521                            & \hspace{1em}547                            & \hspace{1em}2.34        & \hspace{1em}LED        \\
            \hspace{1em}639.5                & \hspace{1em}639                            & \hspace{1em}640                            & \hspace{1em}1.94        & \hspace{1em}Laser      \\
            \hspace{1em}660                  & \hspace{1em}651                            & \hspace{1em}667                            & \hspace{1em}1.88        & \hspace{1em}LED        \\
            \hspace{1em}787                  & \hspace{1em}786                            & \hspace{1em}788                            & \hspace{1em}1.58        & \hspace{1em}Laser      \\
            \hspace{1em}950                  & \hspace{1em}905                            & \hspace{1em}970                            & \hspace{1em}1.31        & \hspace{1em}LED        \\
            \hspace{1em}972                  & \hspace{1em}970                            & \hspace{1em}973                            & \hspace{1em}1.28        & \hspace{1em}Laser      \\
            \hspace{1em}1028                 & \hspace{1em}1027                           & \hspace{1em}1029                           & \hspace{1em}1.21        & \hspace{1em}Laser      \\
            \bottomrule
        \end{tabularx}
    \caption{In order to measure the photoelectric absorption cross section over a range of energies, a series of diodes with known wavelengths was used. The transmission curves provided by the manufacturer are summarized in the peak, lower half-max (LHM), and upper half-max (UHM) columns. They were controlled with a Thorlabs DC2200 LED Driver.}
    \label{tab:Diodes}
\end{table}

To ensure proper alignment between the beam spot and the center of each hole, first a rough x-y value for each hole center was estimated, then a 14-by-14 grid of x-y positions was made for each hole, covering a \SI{0.6}{\milli\meter}-by-\SI{0.6}{\milli\meter} area surrounding that hole's estimated center. Measurements were then taken at each point in these grids, in order that the maximum values for the charge collection, corresponding to precise alignment of the beam spot with a hole, could be used for the analysis.

Furthermore, to eliminate any temporal effects from biasing the results (such as a small increase in fridge temperature over the course of the run), rather than performing the scan over each hole one after another in sequence, the x-y points from all the separate scans were combined and shuffled, so that the order of measurements favored no hole in particular.

To reduce the effect of charge buildup in the crystal over time, the crystal was grounded between every measurement. While grounded, the mirror was used to direct the beam spot toward the open hole, and a \SI{1}{\milli\second} flash of light was sent into the crystal to aid in the neutralization of the charge buildup. Then, the voltage was reapplied to the crystal so the next measurement could be taken.

\section{Calibration}\label{sec:Calibration}

Since each measurement involved a train of 32 light pulses, each voltage trace recorded by the DAQ was split into 32 sections and summed together to form an average pulse. The amplitude of the average pulse was used as measure of the total amount of light transmitted through the silicon and collected by the crystal during that measurement. 

In order to compare two different pulse amplitudes, an absolute calibration of the detector response was needed. This was done in two steps. First, a diode was connected to the system, and the mirror directed the beam spot toward the open hole. Then, a sweep over applied current to the diode was performed, and the average pulse size at each value of the applied current was observed and recorded. Second, the same diode was placed in a dark box with a photomultiplier tube capable of counting single photoelectrons. Using the same sweep over applied current as before, an absolute measure of the crystal response to a given amount of light was obtained. What was found was that the crystal responds linearly with incident light up until average voltage pulses of \SI{2}{\volt}, after which it begins to saturate. In the majority of runs, the width of the laser pulses was tuned so that the pulses sizes remained within this linear range, but in some cases the average pulse amplitude extended outside this range, and the absolute calibration was used to correct that amplitude.

\begin{figure}[t]
    \centering
    \includegraphics[width=0.5\textwidth]{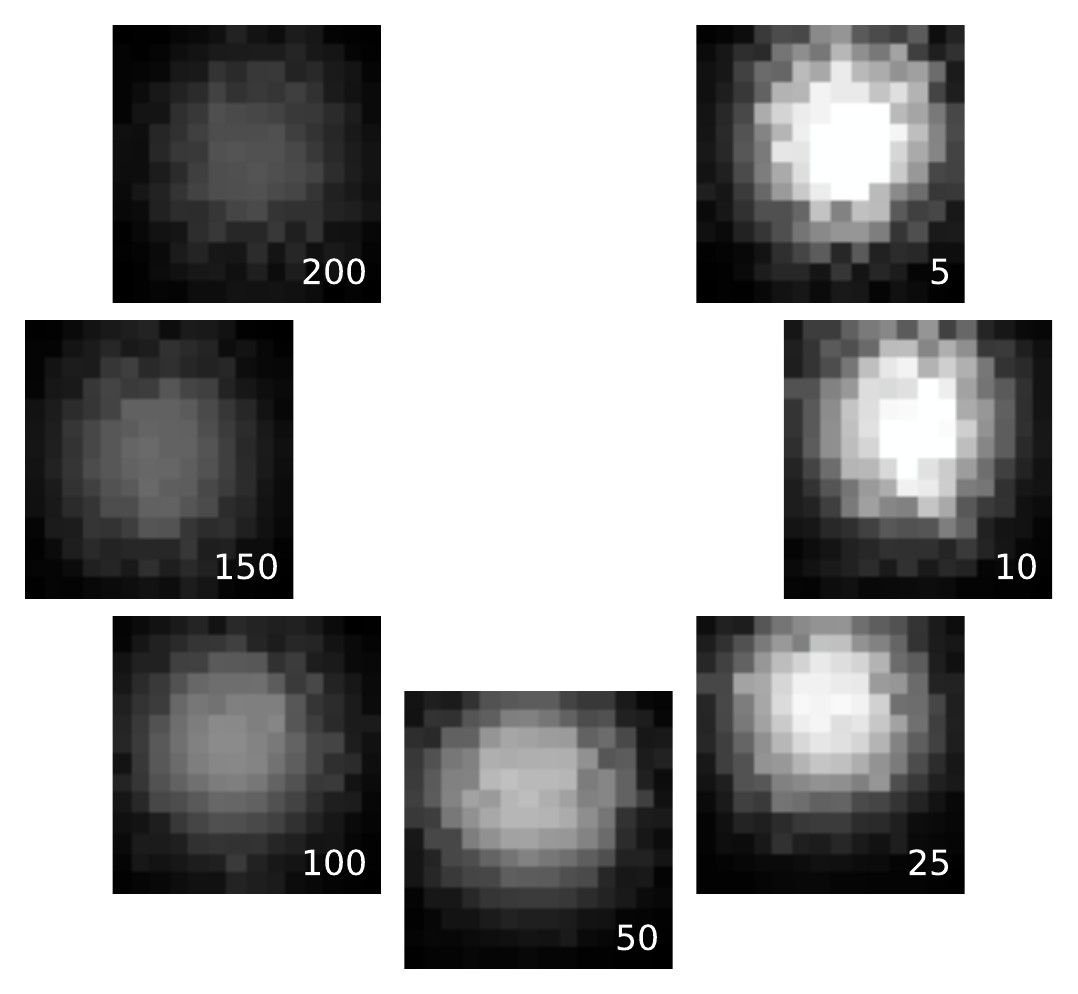}
    \caption{A run using the \SI{950}{\nano\meter} diode at \SI{0.5}{\kelvin}. The run involved a scan over each hole containing a silicon filter. Each scan involved a set of 14-by-14 measurements so that the values corresponding to a precise alignment of the beam spot and the hole could be used for the analysis. The white text indicates the thickness of the silicon filter (in \SI{}{\micro\meter}) corresponding to each scan.}
    \label{fig:GridScan}
\end{figure}

The average amplitudes from all the measurements in each scan were then used to produce a 2-D image of that scan, to visually confirm that the scan was in fact aligned with a hole. The set of scan images for one of the runs can be seen in Figure~\ref{fig:GridScan}, where a decrease in amplitude is seen as the the thickness of silicon increases.

\section{Analysis}\label{sec:Analysis}

To turn a set of scans into a value for the photoelectric absorption cross section, a series of post-processing steps was performed.

First, the set of 14-by-14 points for each scan was divided into thirds based on time of acquisition, representing the first third of points taken for that scan, the second third, and the final third. Since the points for all the scans were taken in a random order, this process effectively separated each scan into 3 ``sub-scans" of lower resolution. 

Second, in order to reduce the impact of outliers, the median of the top 5 amplitudes for each sub-scan was taken as the transmission value for that sub-scan. Finally, these three sub-scan transmission values were taken together, and the median of those values was taken as the transmission value for the whole scan. This reduced the impact of any transient effects at the start of a run that were present in only the first sub-scan.

\begin{figure}[t]
    \centering
    \includegraphics[width=0.5\textwidth]{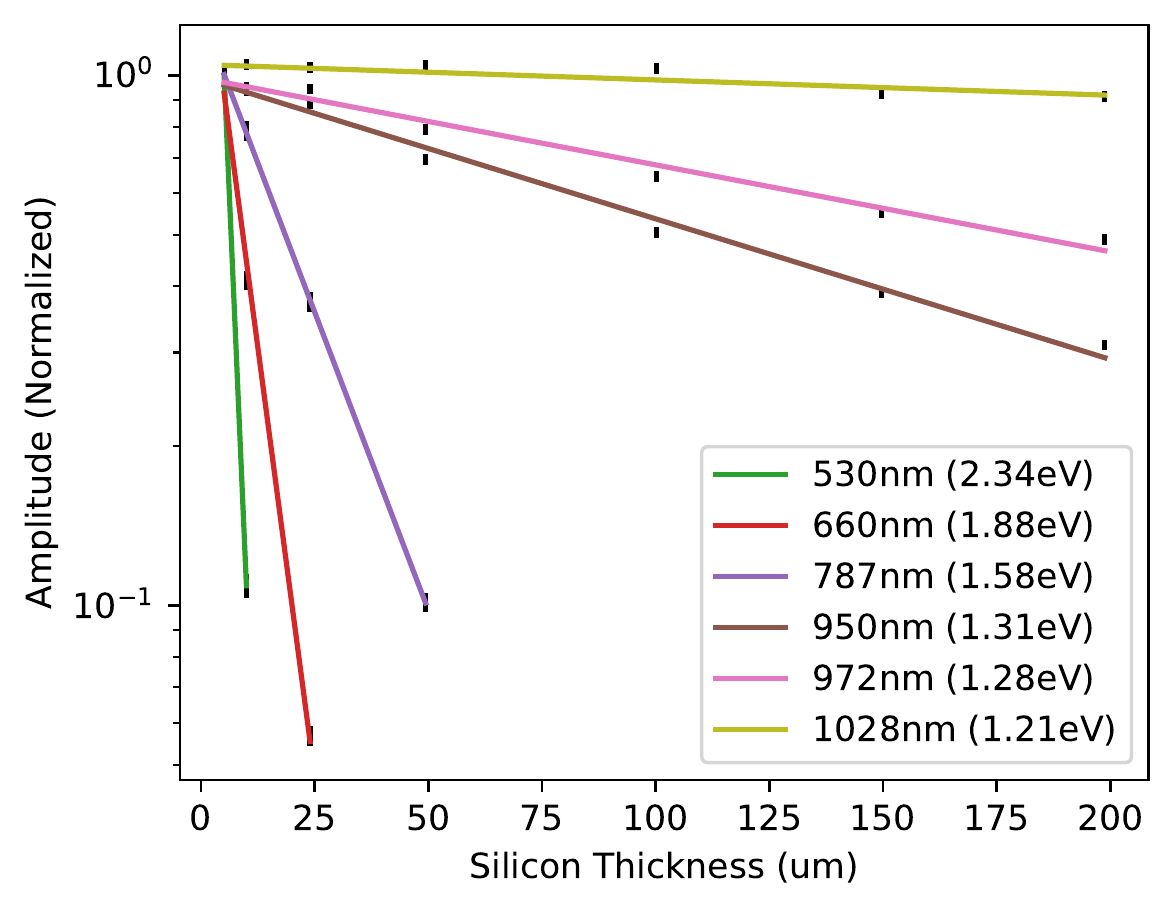}
    \includegraphics[width=0.5\textwidth]{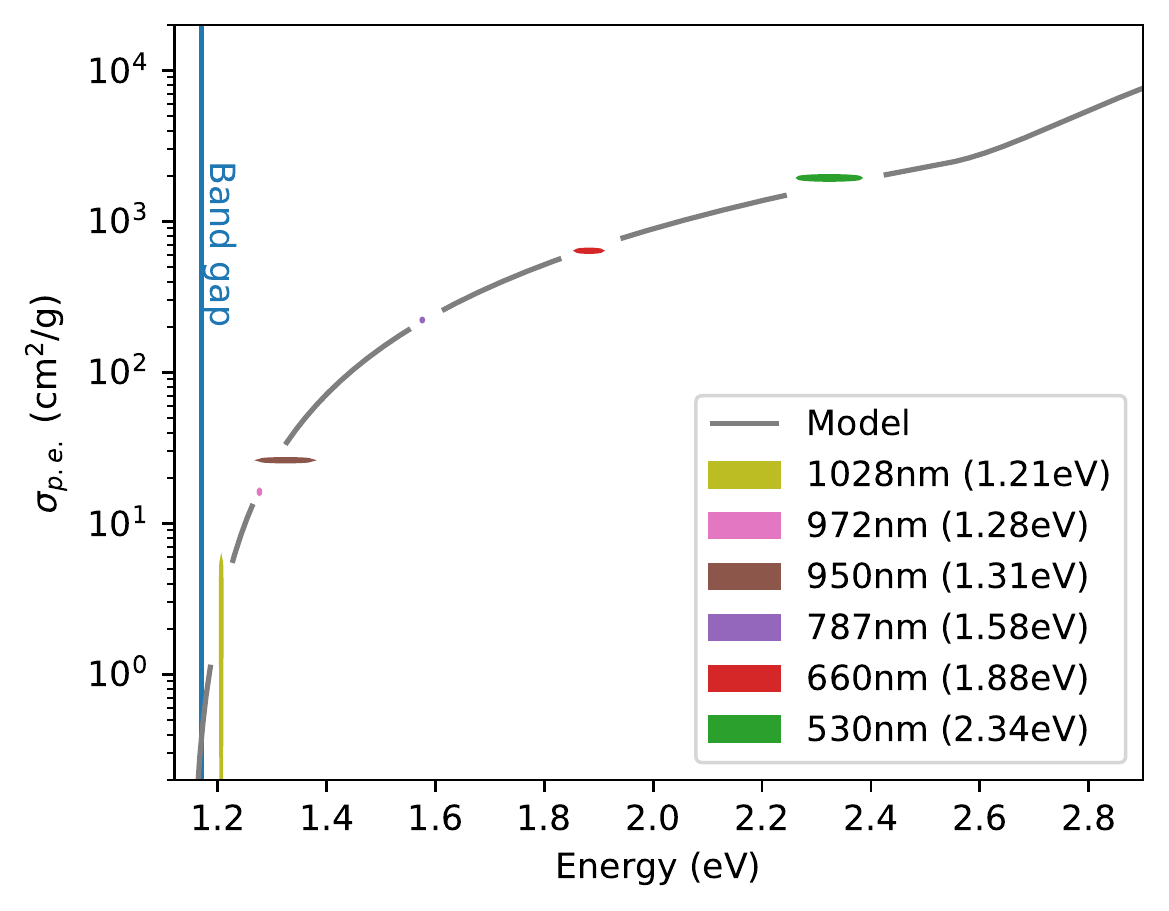}
    \caption{\textbf{Top}: The normalized transmission through the different thicknesses of silicon for multiple wavelengths of light at \SI{0.5}{\kelvin}. The error bars are dominated by systematic errors introduced by the filter mount. The negative of each fitted slope gives the value of the photoelectric absorption cross section at the corresponding wavelength. \textbf{Bottom}: The values of \sigmape\ at \SI{0.5}{\kelvin} from above, plotted in energy space. Each fitted value is presented as an ellipse, with the height representing the uncertainty of the fit and the width representing the uncertainty in the diode wavelength.}
    \label{fig:Fits}
\end{figure}

Once these transmission values were obtained, they were plotted against the corresponding silicon thickness. The results from several runs taken at \SI{0.5}{\kelvin} are shown in Figure~\ref{fig:Fits} (Top). Note that some runs include more points than others. This is because the shorter wavelength light used for some of the runs could not be detected through the thicker pieces of silicon.

The error bars for these points were the result of a detailed study, which checked for biases introduced by various effects, such as bias voltage, light intensity, linearity correction, neutralization procedure, and day-to-day variability. However, due to the symmetric nature of the data-taking process, these were all found to be sub-dominant to the primary source of error, which was caused by the filter mount itself. This was determined by performing a calibration run where each hole of the filter mount was left empty. The standard deviation in the light collection through the eight holes in this case was measured to be approximately 3\%, with a small dependence on wavelength. A run with another mount produced by the same machining process resulted in a different variation from hole to hole, but with the same overall magnitude of variation. Since this was the dominant source of error, the standard deviation of light collection found in this calibration run was used to set the error bars in Figure~\ref{fig:Fits}~(Top).

After plotting, the points were fit with an exponential according to the Beer-Lambert Law:

\begin{equation}
    Ae^{-\rho\sigma_{\mathrm{p.e.}} x}
\end{equation}

\noindent where $A$ (arbitrary constant) and \sigmape\ are fit parameters, $\rho$ is the density of silicon (\SI{2.33}{\gram\per\centi\meter\cubed}), and $x$ is the silicon thickness.

These fitted values for \sigmape\ at \SI{0.5}{\kelvin} are shown in Figure~\ref{fig:Fits}~(Bottom). Each fitted value is presented as an ellipse, with the height representing the statistical uncertainty in the fit, and the width representing the uncertainty of the diode energy. For the energy uncertainty, the lower and upper bounds were calculated from the upper half-max wavelength and lower half-max wavelength of the emission spectrum, respectively (see Table~\ref{tab:Diodes}). 

\section{Temperature Dependence}

\begin{figure}[h!]
    \centering
    \includegraphics[width=0.485\textwidth]{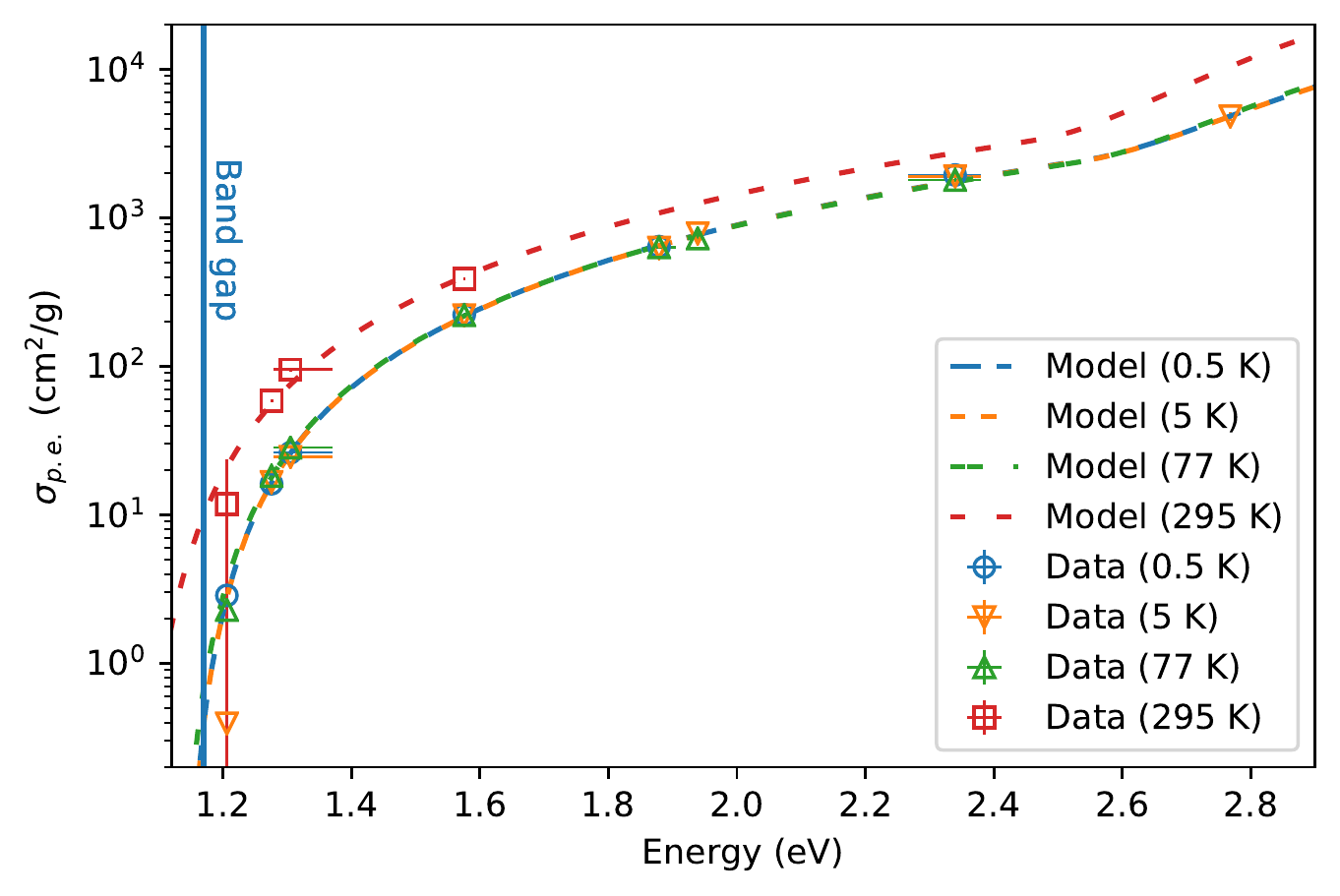}
    \includegraphics[width=0.485\textwidth]{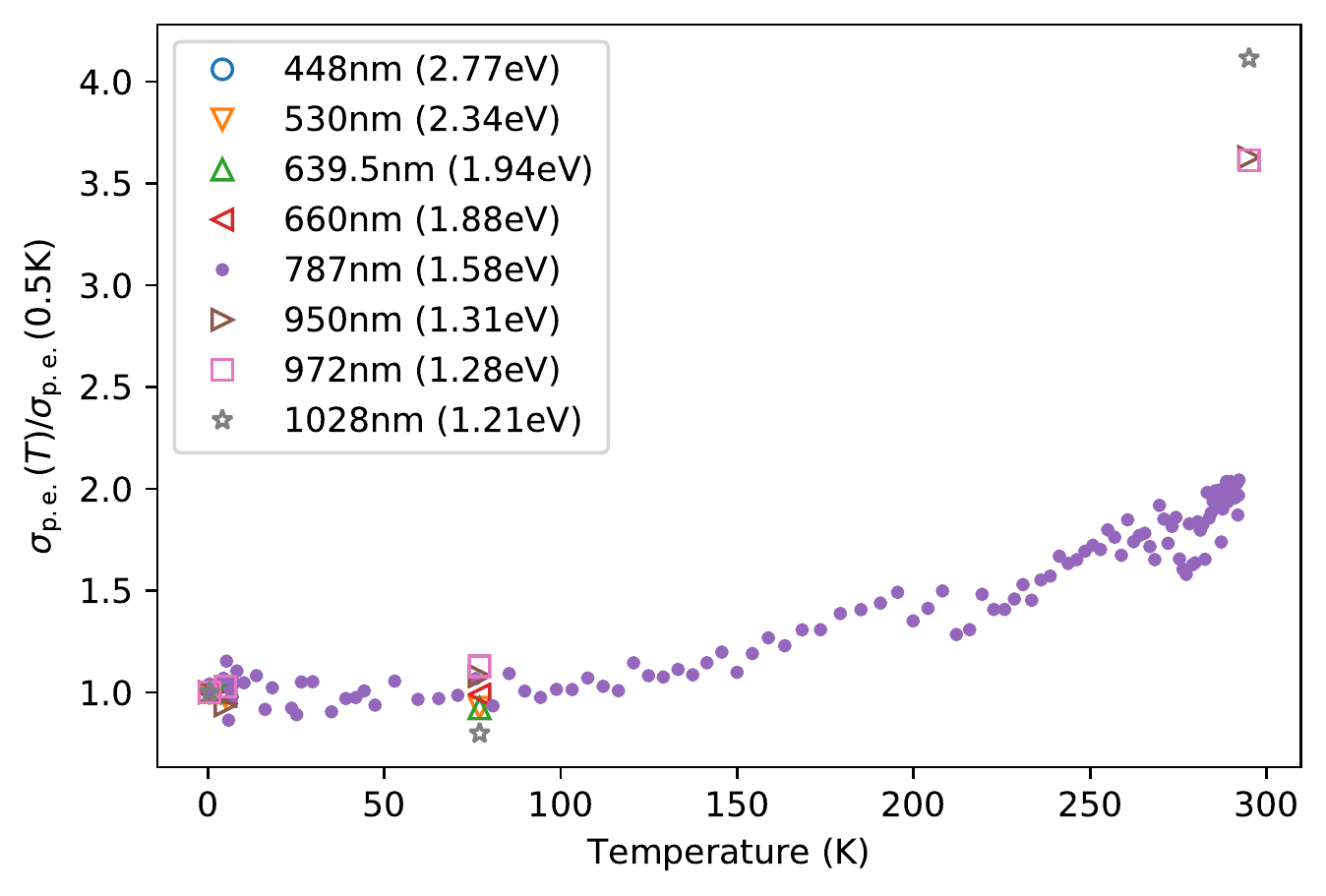}
    \includegraphics[width=0.485\textwidth]{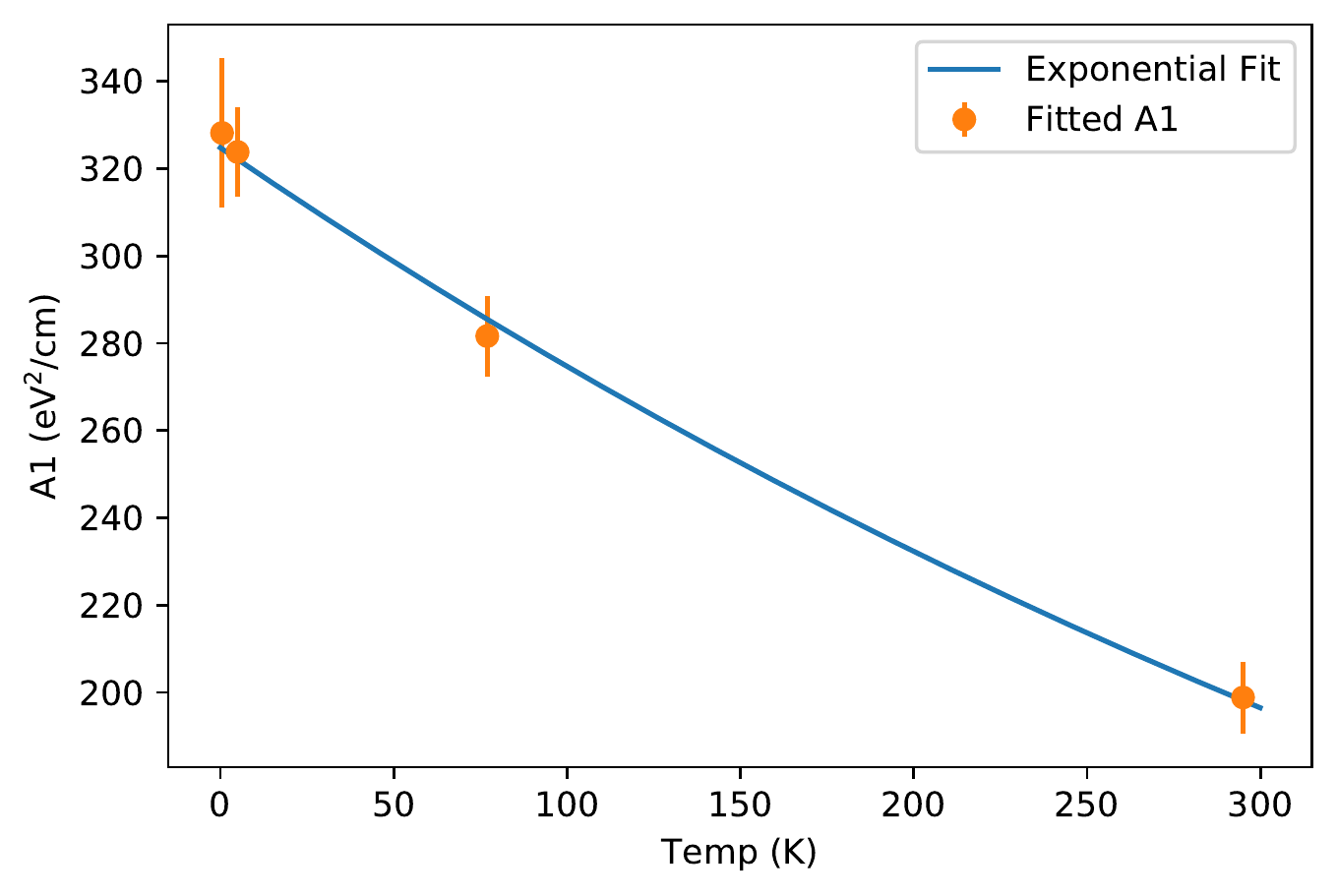}
    \caption{\textbf{Top}: Measurements of \sigmape\ at four discrete temperatures, and the results of a simultaneous fit across all four temperatures using the model described in the text. \textbf{Middle}: The temperature dependence of \sigmape\ relative to $T=\SI{0.5}{\kelvin}$. The cross section for \SI{787}{\nano\meter} was measured continuously as the fridge warmed up, while the cross sections for the other wavelengths were only measured at four discrete temperatures. We did not measure a statistically significant difference in the cross section for temperatures of \SI{77}{\kelvin} and below. \textbf{Bottom}: When the model includes an independent parameter for the first indirect band gap proportionality constant ($A_1$) at each of the four measured temperatures, the fitted values follow an exponential curve.}
    \label{fig:TempDep}
    
\end{figure}
The photoelectric absorption cross section is not expected to differ significantly between \SI{0}{\kelvin} and the \SI{0.5}{\kelvin} results presented in this paper. However, it does increase at warmer temperatures. We confirmed this effect by repeating the \SI{0.5}{\kelvin} measurements at \SI{5}{\kelvin}, \SI{77}{\kelvin}, and \SI{295}{\kelvin}. We also performed some continuous measurements at a fixed wavelength as the fridge was warming up. These data are summarized in Figure~\ref{fig:TempDep}. We did not measure a statistically significant difference in \sigmape\ for temperatures up to \SI{77}{\kelvin}, but did observe an upward trend in \sigmape\ between \SI{77}{\kelvin} and \SI{295}{\kelvin}.

\section{Applicability to dark matter searches}

To set a dark matter limit, a continuous curve in \sigmape–energy space is preferred. To obtain this curve, we fit the discrete measurements to a model describing the temperature-dependent absorption coefficient $\alpha(T)$ via direct and indirect photon absorption~\cite{Rajkanan:1979}:
\begin{multline} \label{eq:abs_model}
    \alpha(T) = \sigma_{\mathrm{p.e.}}(T)\rho_{\textrm{Si}} = \sum_{i,j = 1,2} C_{i}A_{j} \Bigg[ \frac{\left( E_{\gamma} - E_{gj}(T) + E_{pi}\right)^{2}}{e^{E_{pi}/kT} - 1} \\ + \frac{\left( E_{\gamma} - E_{gj}(T) - E_{pi}\right)^{2}}{1-e^{-E_{pi}/kT}} \Bigg] + A_{d}\left( E_{\gamma} - E_{gd}(T) \right)^{1/2}
\end{multline}
where $E_{\gamma}$ is the photon energy, $k$ is the Boltzmann constant, the first and second terms in the sum describe indirect photon absorption via phonon absorption and emission, respectively, and the last term describes direct photon absorption. The suffix $i$ refers to the two phonon energies from the transverse acoustic ($TA$) ($E_{p} = 18.27$\,meV) and transverse optical ($TO$) ($E_{p} = 57.73$\,meV) lattice waves considered, and the suffix $j$ refers to the two different indirect band gaps $E_{g}$ which may be active in phonon absorption. 
We did not explicitly include in the model any direct contributions from possible $B_{s}O_{2}$ complexes~\cite{markevich19, hornbeck55} in the CZ-grown silicon filters.
The $C_{i}$ coefficients describe the electron-phonon coupling constant; for silicon, $C_{TA} = 5.5$ and $C_{TO} = 4.0$. The $A_{1}$, $A_{2}$, and $A_{d}$ coefficients are used as proportionality constants.

The temperature-dependent indirect band gap energies $E_{gj}(T)$ are given by:
\begin{equation}
    E_{gj}(T) = E_{gj}(0) - \frac{\beta T^{2}}{T+\gamma},
\end{equation}
where $\beta = \SI{7.021e-4}{\electronvolt\per\kelvin}$ and $\gamma = \SI{1108}{\kelvin}$. To improve the fit of the model to the lower energy measurements, we allowed the lowest band gap energy at \SI{0}{\kelvin} ($E_{g1}(0)$) to float while using the fixed value of $E_{g2}(0) = \SI{2.5}{\electronvolt}$ for the second indirect band gap energy, where there is not enough data to constrain.

The model was fit simultaneously to the four \sigmape\ measurements taken at 0.5, 5, 77, and 295\,K, taking into account the uncertainties in the measured \sigmape\ values as well as the uncertainties due to the diode wavelength distributions (see Table~\ref{tab:Diodes}). In this fit, $E_{g1}(0)$ was held constant across temperatures, but in order to investigate a possible temperature dependence in $A_{1}$, we allowed $A_{1}$ to vary. Only one measurement was taken at an energy above the second indirect band gap, so the $A_{2}$ temperature dependence was not investigated. 

The results of the fit for $A_{2}$ and $E_{g1}(0)$ are \SI{6(3)e3}{\electronvolt\squared\per\centi\meter} and \SI{1.134(4)}{\electronvolt}, respectively. The fit results for $A_1$ at each of the four temperatures are shown in Figure~\ref{fig:TempDep} (Bottom), and demonstrate a significant temperature dependence. This may be a result of temperature-dependent effects that the phenomenological absorption model described in Equation~\ref{eq:abs_model} does not account for, such as the effect of temperature on the density of states and the electron-phonon coupling. In order to improve the model's fit to the \sigmape\ measurements and produce a result that can be used for dark matter searches, we performed a second iteration of the simultaneous fit, this time constraining $A_1$ to have an exponential temperature dependence:
\begin{equation}
A_{1}(T) = c_{0}e^{-c_{1}T}.
\end{equation}
Using the previously determined $E_{g1}(0)$ and $A_{2}$ as fixed parameters, the fitted $c_{0}$ and $c_{1}$ values are:
\begin{equation} \label{eq:fits2}
\begin{split}
    c_{0} &= 325(6) \,\textrm{eV}^{-2}\textrm{cm}^{-1} \\
    c_{1} &= 1.7(1) \times 10^{-3} \,\textrm{K}^{-1}.
\end{split}
\end{equation}
The result of this fit is shown in Figure~\ref{fig:TempDep}. We chose an exponential function for empirical reasons, as it was physically plausible and required few variables to adequately fit the $A_1$ values. 

The result of this second iteration of the simultaneous fit is shown compared to the data in Figure~\ref{fig:Fits} (Bottom) and Figure~\ref{fig:TempDep} (Top).

\section{Discussion}

The requirement that the source be pulsed for this measurement restricted the light sources primarily to single-wavelength diodes, as tunable light sources are generally continuous. This limitation was largely a function of the charge amplifier, which is AC-coupled, designed for fast charge transport measurements rather than integrated power. Finer sampling of the photo-electric cross-section may be accomplished in future work by modifying the readout to accommodate a filtered xenon flash lamp (TDS) or by modifying the integrator to work with a longer time-constant tunable light source with a conventional shutter, with switching times on the scale of milliseconds.

There is an active interest in using alternative semiconductors in dark matter searches \cite{Hochberg_2018,Griffin_2020,SuperCDMSSoudan2020,Kurinsky_2019,griffin2020sic}. The technique we developed can be easily extended to such materials by placing samples of that material in the filter mount. Note however that to probe the values of the cross section at photon energies below the silicon band gap, a detector crystal with a smaller gap, such as germanium, would need to be used.

\section{Acknowledgements}
We would like to thank Yonit Hochberg for initial conversations that led to the design of this experiment, Brian Lenardo for his assistance with the dark box calibration, and Steve Yellin for helpful comments. This work was supported in part by the U.S. Department of Energy, the National Science Foundation, the DFG (Germany) - Project No. 420484612, and Germany’s Excellence Strategy - EXC 2121 ``Quantum Universe" – 390833306. The fiber feedthrough and LEDs were provided by NSERC Canada. Precision thickness measurements of the silicon filters were provided by Filmetrics (KLA) Application Lab in Santa Clara, CA. This document was prepared by using resources of the Fermi National Accelerator Laboratory (Fermilab), a U.S. Department of Energy, Office of Science, HEP User Facility. Fermilab is managed by Fermi Research Alliance, LLC (FRA), acting under Contract No. DE-AC02-07CH11359. This document was also prepared using resources of SLAC, which is operated under Contract No. DEAC02-76SF00515 with the U.S. Department of Energy.

\bibliographystyle{aipnum4-1}
\bibliography{APL_Si_Transport}

\begin{thebibliography}{30}
\bibitem [{ele()}]{electrode}%
  \BibitemOpen
  \href@noop {} {}\bibinfo {note} {Electrode was a tri-layer of 40nm of W on
  20nm of Al on 40nm of amorphous Si.}\BibitemShut {Stop}%
\end{thebibliography}

\end{document}